\documentclass[a4paper,11pt]{article}
\pdfoutput=1 

\usepackage{jinstpub} 
\usepackage{subfigure}

\title{\boldmath Particle Flow Oriented Electromagnetic Calorimeter Optimization for the Circular Electron Positron Collider}

\author[a,b,c]{H.Zhao}
\author[a]{C.Fu}
\author[d,a]{D.Yu}
\author[a]{Z.Wang}
\author[a]{T.Hu}
\author[a,1]{M.Ruan\note{Corresponding author.}}

\affiliation[a]{Institute of High Energy Physics, Chinese Academy of Sciences\\Beijing 100049, China}
\affiliation[b]{CAS Center for Excellence in Particle Physics}
\affiliation[c]{Collaborative Innovation Center for Particles and Interactions}
\affiliation[d]{Laboratoire Leprince-Ringuet, Ecole Polytechnique\\Palaiseau, France}

\emailAdd{manqi.ruan@ihep.ac.cn}

\abstract{The design and optimization of the Electromagnetic Calorimeter (ECAL) are crucial for the Circular Electron Positron Collider (CEPC) project, a proposed future Higgs/Z factory. Following the reference design of the International Large Detector (ILD), a set of silicon-tungsten sampling ECAL geometries are implemented into the Geant4 simulation, whose performance is then scanned using Arbor algorithm. The photon energy response at different ECAL longitudinal structures is analyzed, and the separation performance between nearby photon showers with different ECAL transverse cell sizes is investigated and parametrized. The overall performance is characterized by a set of physics benchmarks, including $\nu\nu H$ events where Higgs boson decays into a pair of photons (EM objects) or gluons (jets) and $Z\to\tau^+\tau^-$ events. Based on these results, we propose an optimized ECAL geometry for the CEPC project.}


\begin{document}
\maketitle
\flushbottom

\section{Introduction}
\label{sec:1}
After the Higgs boson discovery~\cite{Higgs1,Higgs2}, precise measurements of the Higgs boson properties and the Standard Model (SM) parameters are essential for the particle physics. 
Proposed by the Chinese high energy physics community, the CEPC~\cite{CEPC} project will take advantage of the clean environment of e$^{+}$e$^{-}$ collisions to increase the precision of Higgs boson properties and Electroweak measurements.
The physics program at CEPC requires a detector that can reconstruct a full spectrum of physics objects with great efficiency and precision. 
The Particle Flow Algorithm (PFA)~\cite{PFA,CMS,ATLAS} oriented detector design is therefore of strong physics interests, as it reconstructs every final state particle, providing remarkable efficiency in physics object reconstruction.

Segmented in both longitudinal and transverse directions, a high granular ECAL is crucial for the PFA as it provides the necessary separations between the shower clusters induced by different final state particles.
On the other hand, such PFA oriented ECAL contains a large number of readout channels, which consumes lots of power and increases the construction cost.
Unlike the ILC which will work on the power pulsed mode, the CEPC will work continuously.
According to the CEPC PreCDR~\cite{CEPC}, the conceptual benchmark SiW ECAL has 10$^{7}$-10$^{8}$ readout channels.
The power consumption of each channel is around 10 mW, which means the total power consumption of the ECAL reaches 10$^{5}$-10$^{6}$ W.
Reducing the readout channel number is crucial to control the power consumption and lower the construction cost. 
 
The main functions of the PFA-oriented ECAL are the reconstruction of the photon and the separation between nearby showers.
Specifically, the critical performance requirements for the ECAL are as follows:
\begin{itemize}
\item	A clear Higgs boson mass distribution should be reconstructed from the $Higgs\to\gamma\gamma$ events, which requires a $\sim$16\%/$\sqrt{E}\oplus$1\% photon energy resolution.
\item	Good jet energy response, which could be characterized by the boson mass resolution with hadronic final states. Analysis indicates that once the boson mass resolution reaches a level better than 4\%, the W, Z and Higgs boson could be clearly separated via their invariant mass.
\item	Separation performance, which is crucial for the jet energy resolution and for physics with taus. The benchmark requirement is to efficiently separate the photons generated at Z$\to\tau\tau$ events at 91.2 GeV center of mass energy.
\end{itemize}
Besides these, the ECAL should also provide enough shower profile information for the particle identification~\cite{LICH}. 

Reducing the number of readout channels means either reducing the number of longitudinal layers or increasing the transverse cell size.
Either of them will surely affect the performance of the ECAL. 
The former will mainly affect the photon energy resolution and lepton/particle identification~\cite{LICH}. 
The latter will cause a degradation in the position and angular measurement for the neutral particles, especially the photons. 
More importantly, increasing the transverse cell size will significantly limit the shower separation performance, which is of key importance for the PFA.

In this manuscript, we will focus on the optimization study of these local ECAL geometry parameters: total absorber thickness, longitudinal layer number, silicon sensor thickness, and transverse cell size. 
Using Arbor~\cite{Arbor} algorithm and starting from the CEPC benchmark detectors, we explore the performance dependence on these parameters at a set of benchmark performance. 
For the longitudinal structure, the parameters are compared through the intrinsic photon energy resolution, which is characterized by single photon energy resolution and the mass resolution of Higgs boson via di-photons final states. 
For the transverse cell size, the EM-shower separation capability is chosen as the benchmark, and the impact is also evaluated on the Higgs boson mass resolution with jet final states ($\nu\nu Higgs \to \nu\nu gluongluon$).

This manuscript is organized as follows. 
Section~\ref{sec:2} gives a brief introduction to the simulated geometries and software tools. 
Section~\ref{sec:3} is devoted to the photon reconstruction and EM-shower separation performance. 
The results of optimization studies are reported in Section~\ref{sec:4} and Section~\ref{sec:5}, including the longitudinal parameters scan and the transverse cell size scan. 
The conclusion is summarized in Section~\ref{sec:6}. 

\section{Geometries and Simulation Tools}
\label{sec:2}
The International Large Detector (ILD)~\cite{ILD} design is the PFA oriented detector design for the future electron positron colliders, providing an excellent reference for the CEPC detector design. 
A conceptual benchmark detector geometry for CEPC Pre-CDR study, CEPC\_v1~\cite{ManqiHGGReport}, was developed and implemented into the Geant4~\cite{Geant4} simulation. 
Similar to ILD, CEPC\_v1 uses sampling structure ECAL that is composed of silicon sensors and tungsten absorber plates. 
The inner radius of the barrel is 1843 mm and the length of the barrel in Z direction is 2450 mm.
The total absorber thickness is 84 mm, which equals 24 X$_{0}$, divided into 30 longitudinal layers. 
Each layer consists of a 500-micrometer silicon sensor layer. 
Transversely, the sensor plate is segmented into 5~mm square readout cells.  
The Moliere radius of the benchmark ECAL is 17.6 mm.
The total detector is installed in 3.5 T magnetic field.

\begin{figure}[htbp]
\centering
\begin{minipage}[t]{0.4\textwidth}
\subfigure{\includegraphics[width=1\textwidth]{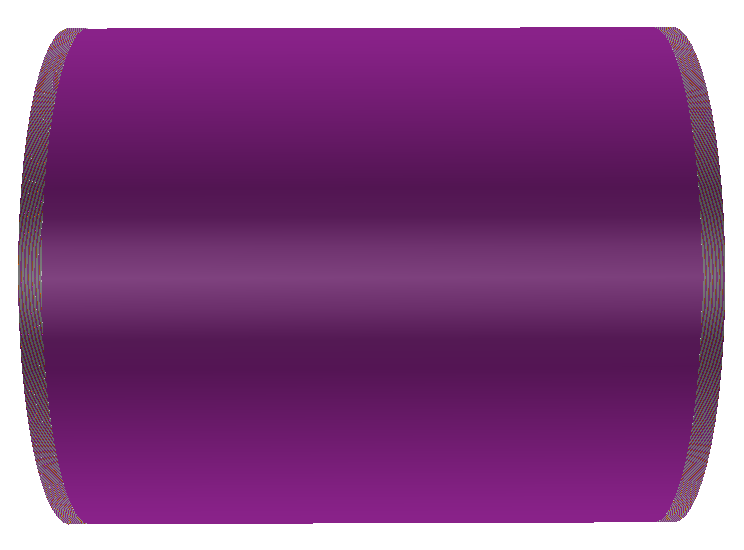}}
\end{minipage}
\begin{minipage}[t]{0.5\textwidth}
\subfigure{\includegraphics[width=1\textwidth]{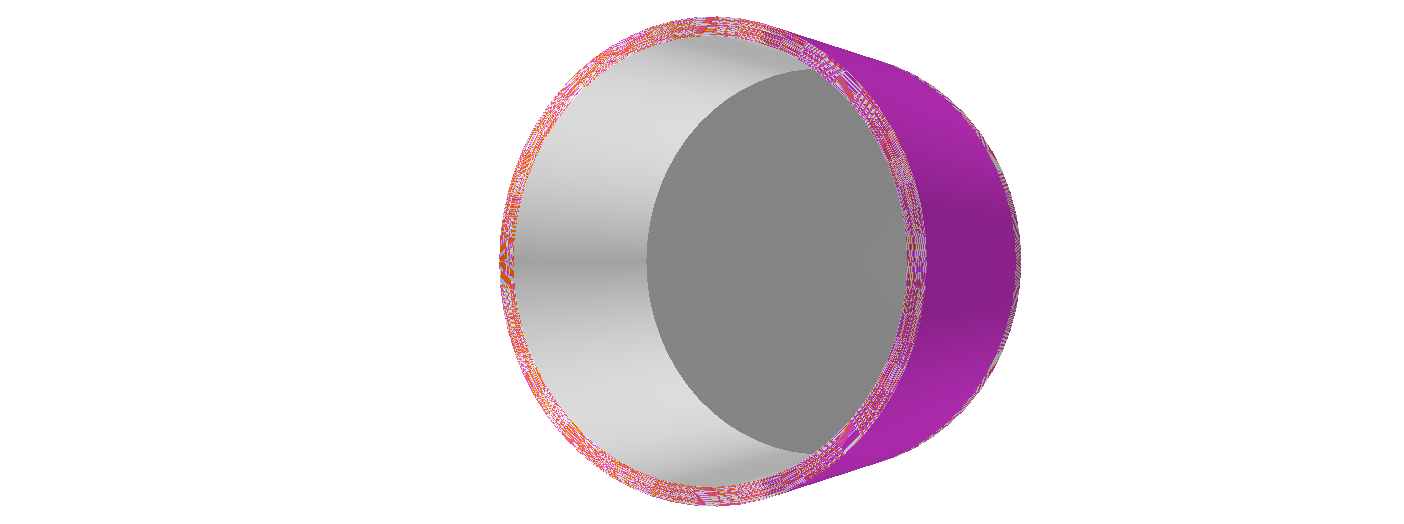}}
\subfigure{\includegraphics[width=1\textwidth]{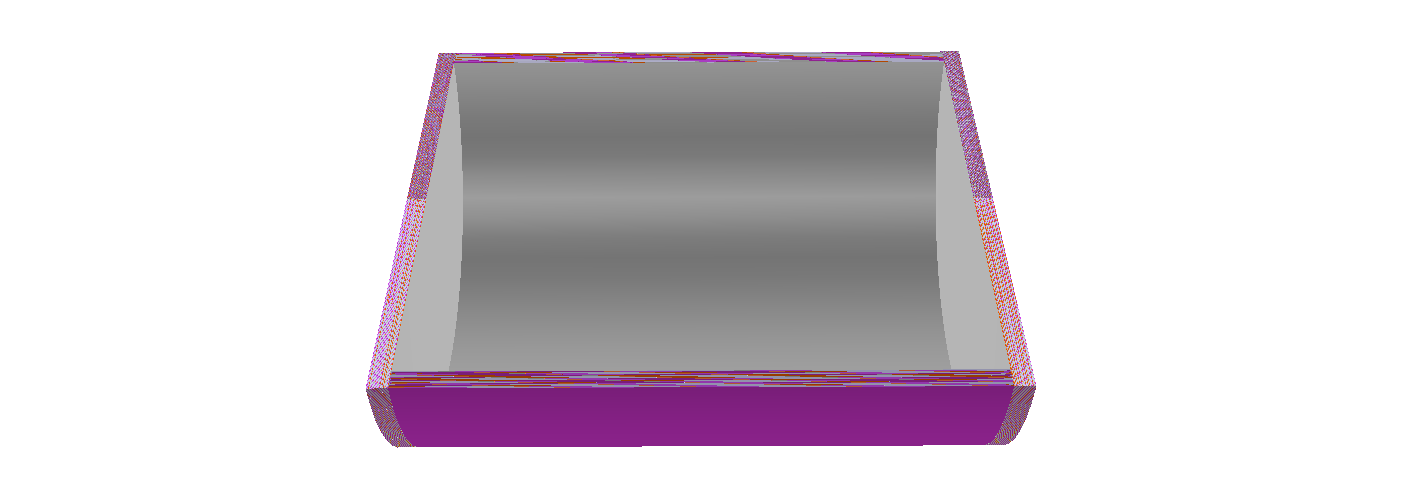}}
\end{minipage}
\caption{\label{fig:1} A simulated simplified ECAL structure.}
\end{figure}

The CEPC\_v1 simulation models the cracks between ECAL modules, staves, and the dead zone between ECAL barrel and endcaps. 
These defects may significantly impact the physics performance. 
Meanwhile,  before reaching ECAL, the photons can also interact with the materials before ECAL. 
To evaluate the impact of geometry defects and the materials before ECAL, a simplified, defect-free ECAL geometry has also been implemented. 
This simplified ECAL geometry consists of a cylindrical barrel and two endcaps, forming a closed cylinder. 
We abstract a few parameters to describe both global structure and local ECAL structure for the simplified ECAL geometry.  
The global parameters include the inner radius, the barrel length, the endcap outer radius, and the opening angles at the endcap. 
For the local structure, the longitudinal layout is described by the number of layers and the material compositions/thickness in each layer, while the transverse structure is described by the cell size in two orthogonal directions. 

In this manuscript, we focus on the local parameters optimization and thus the global parameters are determined according to CEPC v$\_$1. 
The baseline geometry at the CEPC v$\_$1 was used as the start point for local parameters optimization. 
For simplicity, all the tungsten plates have the same thickness.
The simplified ECAL is also divided into 30 layers, each consist of 2.8 mm tungsten, 0.5 mm silicon, and 2 mm PCB layers. 
The transverse cell size is 5 mm by 5 mm.

\section{Photon Reconstruction Using Arbor}
\label{sec:3}
PFA attempts to identify and reconstruct all the final-state particles in the most suited sub-detector systems. 
Explicitly, PFA reconstructs charged particles in the tracking system, photons in the ECAL and neutral hadrons in the whole calorimeter. 
In this paper, Arbor~\cite{Arbor} has been used for the PFA reconstruction.
Inspired by the fact that the particle shower takes naturally the tree configuration, Arbor connects all the calorimeter hits into tree structure clusters. 
In the ideal case, each cluster corresponds to one final state particle. 
Arbor takes in all the calorimeter hits and reconstructed tracks as input, and reconstructs all the final state particles accordingly.

The PFA performance can be characterized by energy response at single particle sample and separation performance at bi-particle sample.
We will discuss both of them with photons, which are most relative to the ECAL. 
At single photon level, we simulate the photon with the energy of 100 MeV - 175 GeV at simplified ECAL geometry.
10000 events are simulated at each energy point.
At bi-photon level, we simulate two parallel 5 GeV photons with their distance ranges from 1 mm to 80 mm.
1000 events are simulated each time.

\subsection{Reconstruction of single photon}
\label{sec:3.1}
In the high granularity calorimeter, a shower is composed of a compact core and a loose halo. 
The core develops along the direction of the incident particle and contains most of the energy.
The halo consists of many low energy clusters and isolated hits induced by secondary particles and contains minority energy of the shower.
Thus the photon reconstruction could be characterized by the energy collection efficiency, defined as the accumulated hit energy in the photon cluster divided by that in all the hits.
Higher energy collection efficiency usually leads to better energy resolution. 
On the other hand, high energy collection efficiency usually leads to long range connection in cluster reconstruction, which may degrade the separation performance.

To characterize the reconstruction performance of single photon, we study the photon finding efficiency and the photon shower energy collection efficiency at different energies. 
The photon finding efficiency is defined as the probability that Arbor can find at least one cluster once the photon is incident to the ECAL fiducial region. 
The energy collection efficiency is defined as the ratio between energies in the clusters and in all the calorimeter hits. 

With simplified ECAL geometry, the finding efficiency reaches 100\% for photons with energy larger than 500 MeV. 
The finding efficiency decreases to 85\% once the photon energy is reduced to 100 MeV. 
The energy collection efficiency is better than 99\% when the photon energy ranges from 1 GeV to 175 GeV. 
When the photon energy is less than 1 GeV, the energy collection efficiency degrades, i.e., the average energy collection efficiency decreases to 75\% for 100 MeV photons. 
Since the simplified ECAL has no material before the calorimeter, it maintains high efficiencies even for low energy photons.

\subsection{Di-photon separation}
\label{sec:3.2}
The shower separation ability is crucial for the PFA reconstruction. 
It is highly appreciated in the jet energy resolution, the $\pi_{0}$ reconstruction, and the $\tau$ reconstruction. 
We characterize the EM-shower separation performance by the reconstruction efficiency of nearby di-photon events. 
The di-photon sample simulates two parallel photons incident into ECAL. 
According to the photon energy distribution at CEPC, the photon energy is set to 5 GeV.

\begin{figure}[htbp]
\centering 
\includegraphics[width=.58\textwidth]{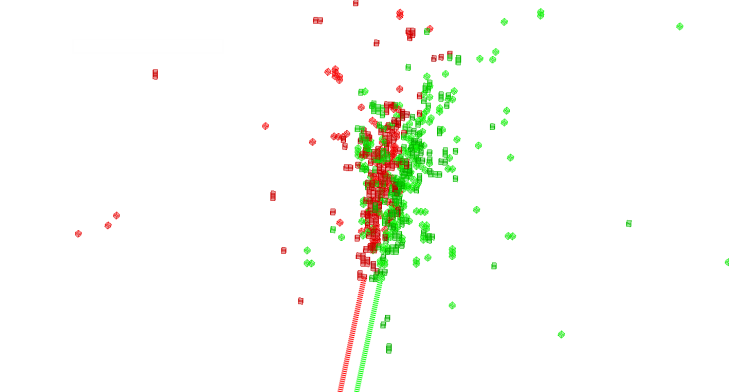}
\qquad
\includegraphics[width=.32\textwidth]{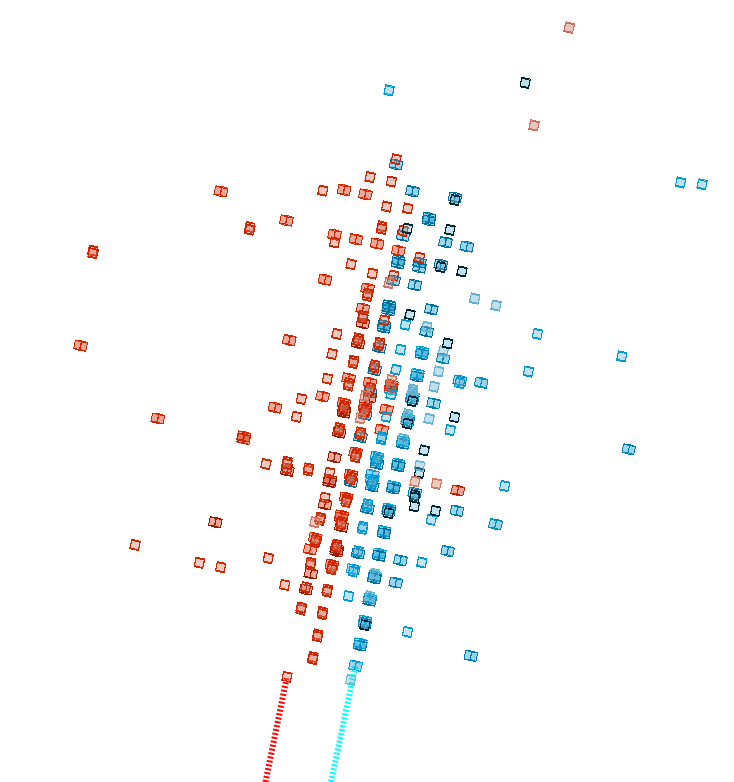}
\caption{\label{fig:3} Reconstructed di-photon events. The left plot is with 1 mm cell size ECAL, each photon has an energy of 5 GeV, and the distance between them is 4 mm. The right plot is with 5 mm cell size ECAL and the distance between them is 11 mm.}
\end{figure}

To quantify the separation performance, we define the reconstruction efficiency as the probability to successfully reconstruct two photons with anticipated energy.
An event would be marked as successfully reconstructed if both the reconstructed photons have more than 1/3 and less than 2/3 of all the deposit energy of the both photons. 
In another word, if one reconstructed photon has more than twice the energy of the other one, the reconstruction will be marked as failed. 
Figure~\ref{fig:3} shows two successfully reconstructed di-photon events with 1 mm and 5 mm cell size ECAL.

\begin{figure}[htbp]
\centering 
\includegraphics[width=.45\textwidth]{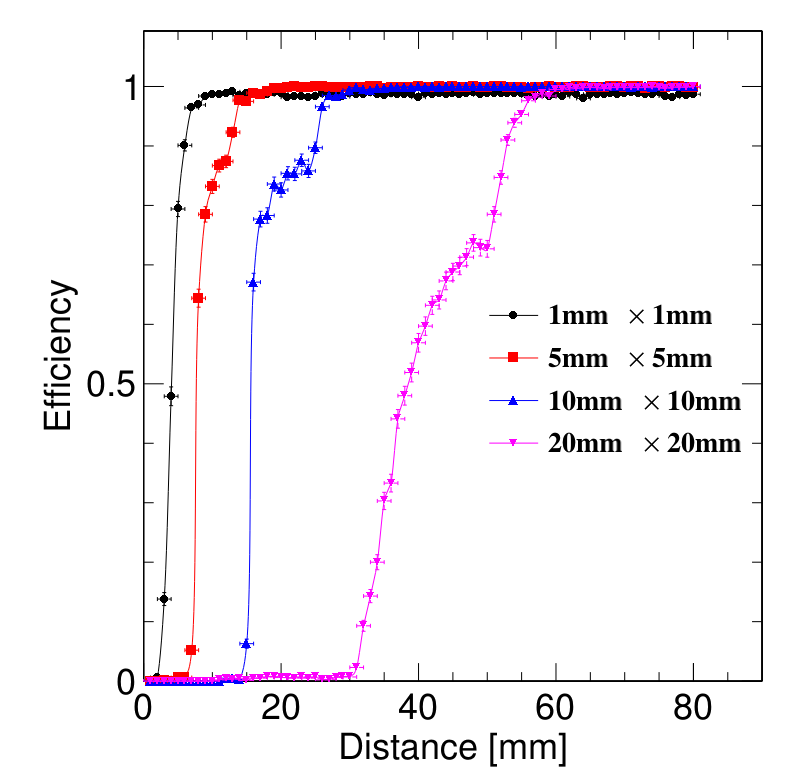}
\caption{\label{fig:4} Reconstruction efficiency of di-photon samples at different cell size. The distance of the photons is scanned from 1 to 80 mm.}
\end{figure}

This probability depends mainly on the separation distance and the ECAL cell size, as shown in Figure~\ref{fig:4}. 
The distance between expected impact points of the two photons is scanned from 1 to 80 mm, with the cell size ranges from 1 mm by 1 mm to 20 mm by 20 mm. 
The parameters of Arbor are adjusted at different cell sizes. 
At large distance, for any cell size, the separation efficiency converges to $\sim$1 since these photon showers become disentangled. 
At very short distance, the showers overlap and the reconstruction efficiency drops steeply.
Once the two photons are shot at exactly the same position, Arbor cannot distinguish the overlapped showers and the reconstruction efficiency vanishes. 
The unsmooth patterns in Figure~\ref{fig:4} are induced by the finite cell size. 

\begin{figure}[htbp]
\centering 
\includegraphics[width=.45\textwidth]{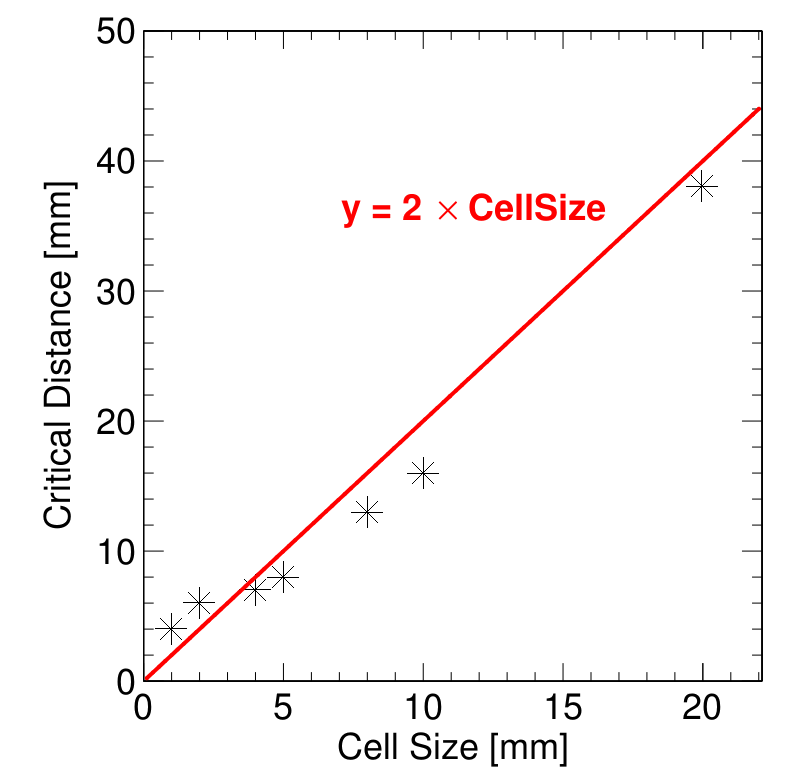}
\caption{\label{fig:5} The critical separation distance at different cell size.}
\end{figure}

The critical separation distance is then defined as the distance at which 50\% of the events are successfully reconstructed. 
When the cell size is larger than 4 mm $\times$ 4 mm and smaller than the Moliere radius, the critical separation distance is roughly twice of the cell size, see Figure~\ref{fig:5}.

\section{Optimization on ECAL Longitudinal Structure}
\label{sec:4}
In CEPC v$\_$1 geometry, the ECAL is longitudinally divided into 30 layers, and each layer contains tungsten as absorber layer, silicon as sensor layer and PCB as electronics/service layer. 
The longitudinal structure parameters are essential for the intrinsic photon energy resolution, .e.g. the total absorber thickness will decide the energy leakage ratio of EM-showers and the thickness of silicon sensor will decide the sampling ratio. 
In this section, we will discuss these effects and propose an optimized longitudinal structure.
We simulate the photon with the energy of 1 GeV - 175 GeV at simplified ECAL geometry with different parameters.
10000 events are simulated each time.

\subsection{Total Absorber Thickness}
\label{sec:4.1}
\begin{figure}[htbp]
\centering 
\includegraphics[width=.45\textwidth]{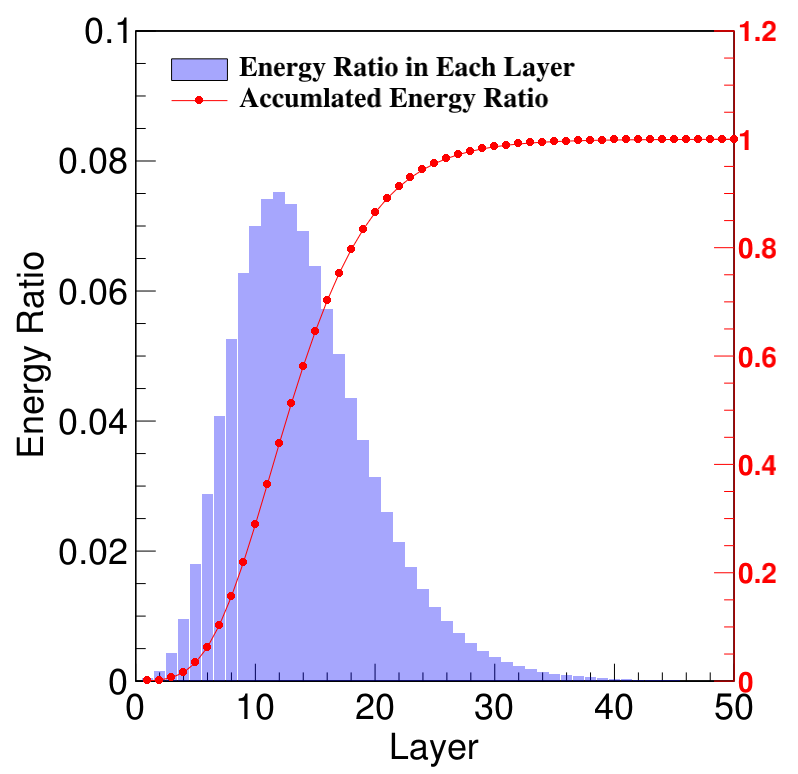}
\caption{\label{fig:6} Ratio of 120 GeV photon shower energy deposited in each layer of SiW ECAL (with 50 layers, 2.8 mm tungsten, 0.5 mm silicon and 2mm PCB in each layer), and the integrated energy before each layer.}
\end{figure}

The total absorber thickness of ECAL can be determined by the requirement on the longitudinal leakage of the most energetic electromagnetic showers. 
Giving the fact that CEPC will be operating at 91-240 GeV center of mass energy, the most energetic electromagnetic particles are generated from the Bhabha events, the ISR return events and the Higgs boson events with Higgs bosons decay into photons. 
The final state photons of $H\rightarrow\gamma\gamma$ decay can be as high as 100 GeV at CEPC. 
And the final state electrons of Bhabha events could reach half of the collision energy. 
Figure~\ref{fig:6} shows the longitudinal energy distribution for 120 GeV photons at a thickened ECAL. 
98.6\% energy of the 120 GeV photon shower deposits at first 30 layers, each has 2.8 mm thick tungsten. 
 
\begin{figure}[htbp]
\centering 
\includegraphics[width=.45\textwidth]{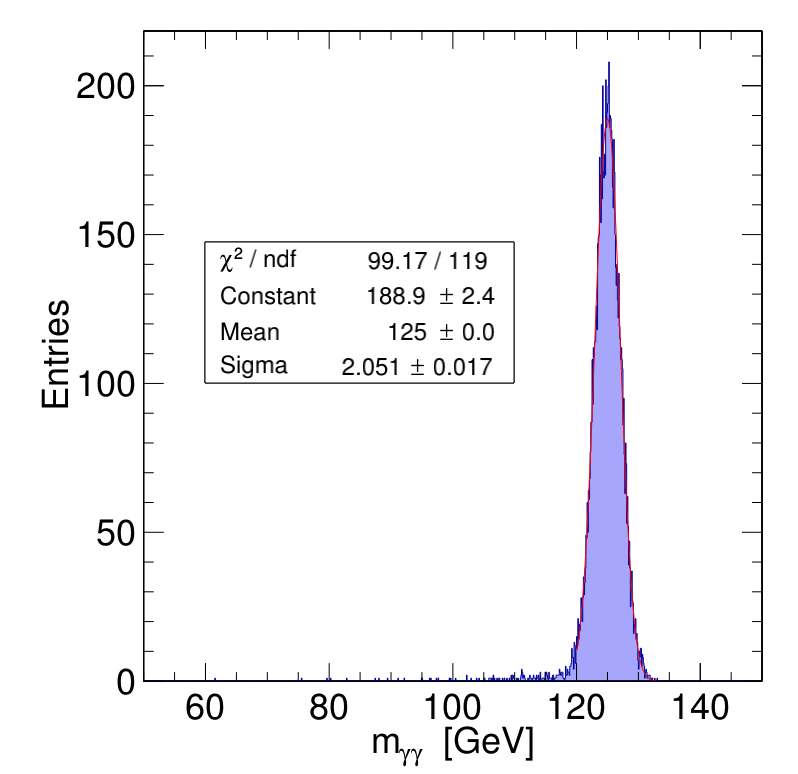}
\qquad
\includegraphics[width=.45\textwidth]{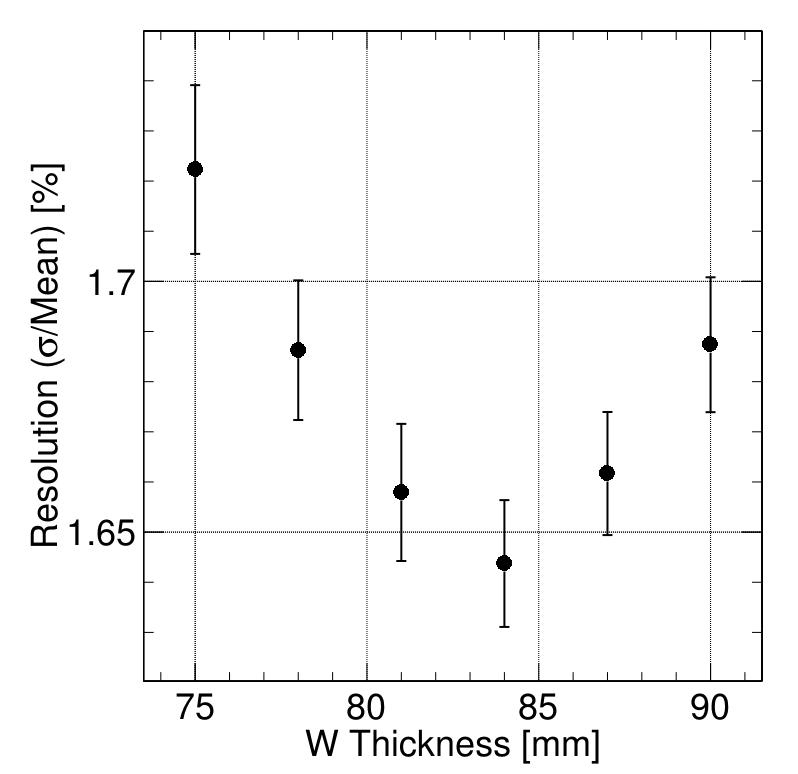}
\caption{\label{fig:7} Higgs boson mass reconstructed from $\nu\nu$Higgs$\rightarrow\gamma\gamma$ events with 84 mm total tungsten thickness (left) and the resolution ($\sigma$/Mean) of reconstructed Higgs boson mass at different tungsten thickness (right).}
\end{figure}

To optimize the total tungsten thickness, we reconstruct the $\nu\nu H, H\rightarrow \gamma\gamma$ events and compare the Higgs boson mass resolution with different absorber thickness. 
At the default setting, a mass resolution of 1.6\% has been achieved, see the left plot of Figure~\ref{fig:7}.
Using thinner absorber, the energy leakage will impact the reconstructed photon energy resolution. 
While at thicker absorber, the resolution is mainly limited by the sampling ratio. 
A scan shows that the best Higgs boson mass resolution is achieved with the default tungsten thickness of 84~mm, see the right plot of Figure~\ref{fig:7}. 

Similar analysis has also been operated at CEPC\_v1 geometry, where the geometry inhomogeneities, especially the cracks between ECAL modules and staves, make a significant impact on the accuracy. 
A geometry based correction, which gives different calibration constants for the photons hit at the geometry defect zones of the ECAL, is mandatory to control those effects. 
After a careful photon energy re-calibration, the degradation induced by these geometry effects can be controlled to better than 20\%, or say, a Higgs boson mass resolution better than 2\% can be achieved with CEPC\_v1 geometry
~\cite{ManqiHGGReport}. 

\subsection{Sensitive Thickness and Number of Layers}
\label{sec:4.2}

\begin{figure}[htbp]
\centering 
\includegraphics[width=.45\textwidth]{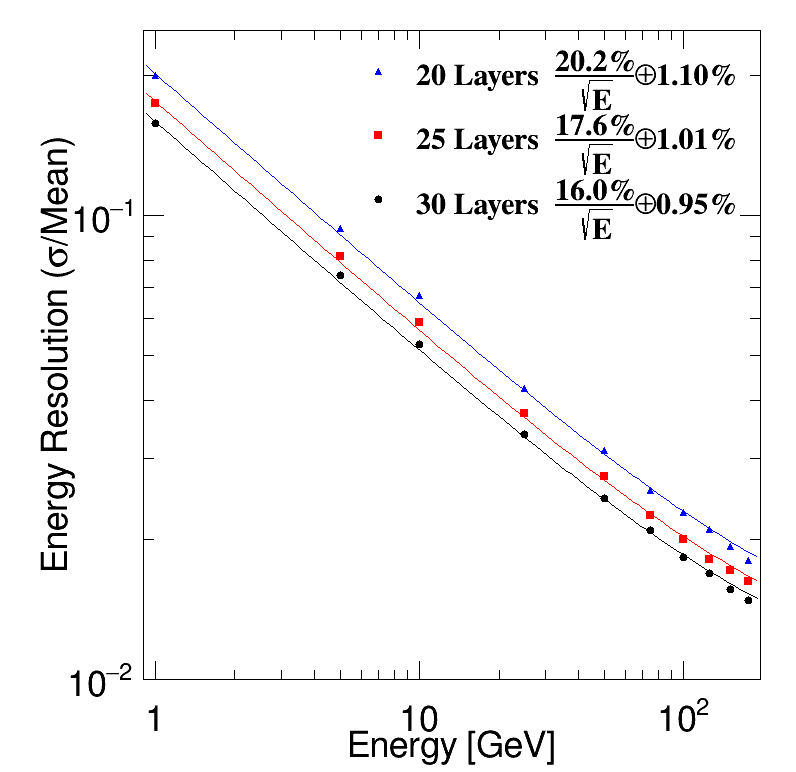}
\caption{\label{fig:8} Impact on photon energy resolution when reducing the number of layers from 30 to 25 and 20, while the total thickness of absorber and the thickness of sensitive layer maintain the same.}
\end{figure}

The single photon energy resolution of the simplified 30-layer ECAL is displayed as the black curve in Figure~\ref{fig:8}, which is consistent with the test beam result of ILD ECAL prototype~\cite{CALICE}.
This set up then serves as a reference for the layer number optimization. 

Reducing the number of layers means fewer read-out channels, which leads to lower construction cost and power consumption. 
Keeping the total absorber thickness at the optimized value of 84~mm, 
reducing the readout layer numbers and maintaining the local sensor thickness, the ECAL energy resolution degrades as the sensor-absorber ratio decreases.
Figure~\ref{fig:8} shows the impact on energy resolution if the number of layers is reduced to 25 and 20. 
Compared with 30 layers option, the energy resolution degrades by 11\% at 25 layers and 26\% at 20 layers. 

\begin{figure}[htbp]
\centering 
\includegraphics[width=.45\textwidth]{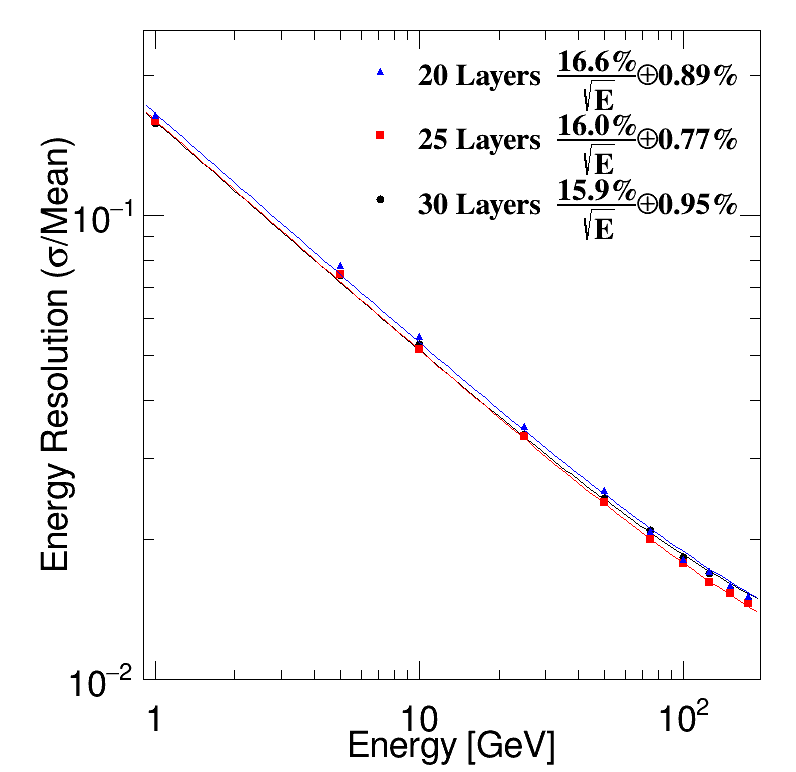}
\caption{\label{fig:9} Energy resolution with less layers and thicker silicon wafer (20 layers with 1.5 mm silicon wafer and 25 layers with 1 mm silicon wafer),  compared to 30 layers and 0.5 mm thick silicon wafer.}
\end{figure}

The degradation of photon energy resolution by reducing the number of channels could be compensated by using thicker silicon sensor. 
We study the energy resolution of ECAL at 20 layers with 1.5 mm thick silicon wafer and 25 layers with 1 mm thick wafer. 
These two options can achieve the same level energy resolution with 30 layers, 0.5 mm thick wafer option, as shown in Figure~\ref{fig:9}. 

\begin{figure}[htbp]
\centering 
\includegraphics[width=.45\textwidth]{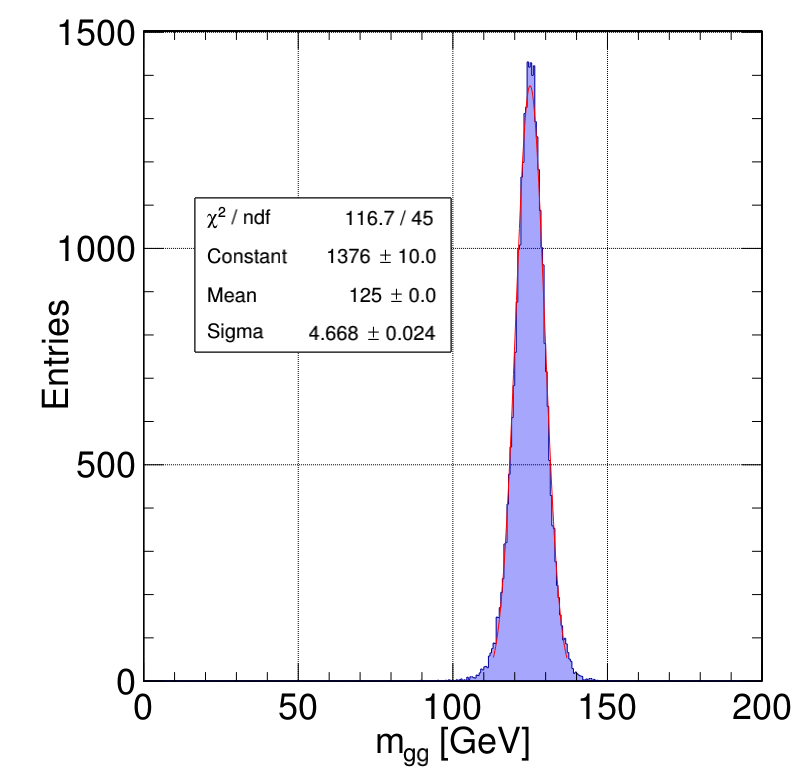}
\caption{\label{fig:10} Higgs boson mass reconstructed from $\nu\nu Higgs, Higgs\to gluons$ events with 30 layers, 0.5 mm silicon ECAL.}
\end{figure}

This conclusion is supported by the reconstruction of jets, which is characterized by the Higgs boson mass resolution at $\nu\nu Higgs, Higgs\to gluons$ events.
As shown in Figure~\ref{fig:10}, using Arbor and the CEPC-v1 geometry, a Higgs boson mass resolution of 3.74\% has been achieved at 30 layers, 0.5 mm silicon ECAL with a standard event selection procedure.
The standard selection has been designed to veto events with energetic and visible ISR photon, events with significant neutrinos generated in gluon fragmentation, and events with jets aiming at beam pipe. 
This event selection has an overall efficiency of 65\%. 
The Higgs boson mass resolutions with other longitudinal structures are shown in Table~\ref{tab:1}.
A marginal difference has been observed in these different configurations. 
In considering possible systematic uncertainties from Arbor parameter optimization at different geometry, we consider these results consistent with the single photon results, which indicates the 20 layers with 1.5 mm thick silicon wafer and 25 layers with 1mm thick wafer options can achieve the same level energy resolution with 30 layers with 0.5 mm thick wafer option.

\begin{table}[htbp]
\centering
\caption{\label{tab:1} Resolution of reconstructed Higgs boson mass through $\nu\nu Higgs, Higgs\to gluons$ events using different longitudinal structures at CEPC\_v1 geometry.}
\smallskip
\begin{tabular}{|c|c|c|}
\hline
Layer number & Silicon sensor thickness & Higgs boson mass resolution (Statistic error only) \\
\hline
30  & 0.5 mm  & 3.74 $\pm$ 0.02 \% \\
25  & 1 mm & 3.71 $\pm$ 0.02 \% \\
20 & 1.5 mm  & 3.78 $\pm$ 0.02 \% \\
\hline
\end{tabular}
\end{table}

\section{Optimization on ECAL Cell Size}
\label{sec:5}

In Section~\ref{sec:3.2}, we discussed the EM-shower separation ability at different ECAL cell size and characterize it with the critical separation distance. 
The impact of separation performance on physics events is evaluated at $Z\to\tau^+\tau^-$ events,  for the following reasons. 
First of all, the reconstruction of physics events with $\tau$ final states leads to rich physics program and is of strong physics interest~\cite{tau}. 
Secondly, the CEPC is a powerful Z factory and can produce hundreds of millions of $Z\to\tau^+\tau^-$ events. 
Thirdly, energetic $\pi_0$s are generated in $\tau$ decay and then decay into very closed photons. 
The successful reconstruction of the photons generated in $\tau$ decay therefore makes a clear physics requirement for the separation performance. 

\begin{figure}[htbp]
\centering 
\includegraphics[width=.5\textwidth]{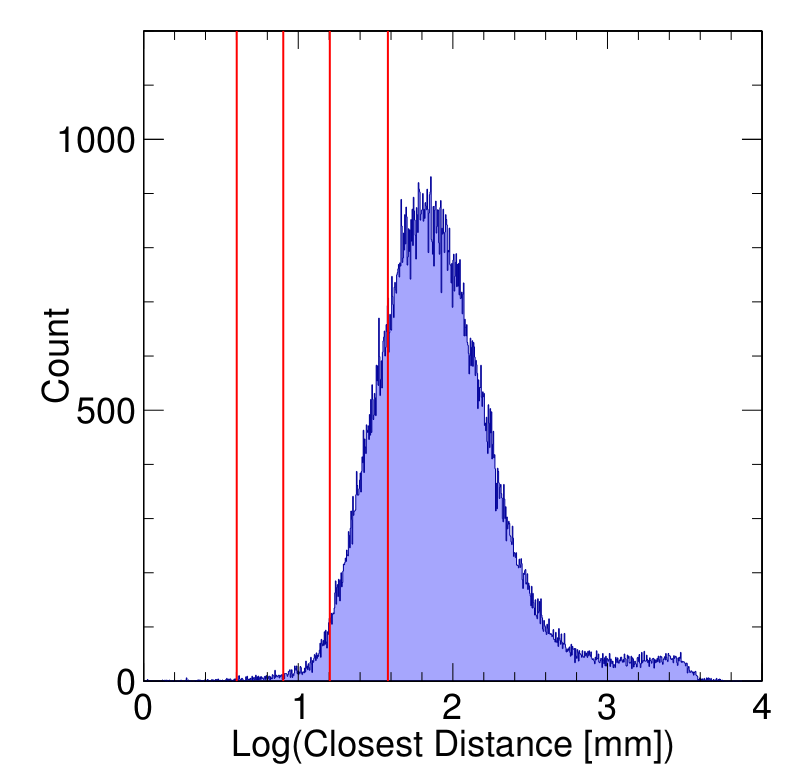}
\caption{\label{fig:11} The distribution of the distance between photon and its closest neighbor, from $Z\rightarrow\tau^+\tau^-$ events at 91.2 GeV. The red line from left to right represents the critical separation distance when cell side length equals 1mm, 5mm, 10mm, 20mm.}
\end{figure}

We calculate the impact position between photon to its closest neighbor(except for neutrino and muon) of the $Z\rightarrow\tau^+\tau^-$ events, as shown in Figure~\ref{fig:11}. 
We derive the percentage of photons whose distance to its closest neighbor is smaller than the critical separation distance with different ECAL cell size.
As shown in Table~\ref{tab:2}, with the ECAL cell size is at 10~mm, the overlapping chance is 1.7\% only. 
However, once the ECAL cell size increases to 20~mm, this overlapping chance rapidly increases by one order of magnitude. 

\begin{table}[htbp]
\centering
\caption{\label{tab:2} Percentages of photons that would be polluted by neighbor particles}
\smallskip
\begin{tabular}{|c|c|c|}
\hline
Cell Size & Critical Separation Distance with Arbor & Percentage of $Z\rightarrow\tau^+\tau^-$ \\
\hline
1~mm  & 4~mm  &0.07\% \\
5~mm  & 8~mm &0.30\% \\
10~mm & 16~mm  &1.70\% \\
20~mm & 38~mm  &19.6\% \\
\hline
\end{tabular}
\end{table}

The reduced separation power also leads to a degraded boson mass resolution. 
Charactered by the Higgs boson mass resolution at $\nu\nu Higgs, Higgs\to gluons$ events, we compare the reconstructed Higgs boson mass resolution at 5~mm/10~mm/20~mm ECAL transverse cell size, as shown in Table~\ref{tab:3}. 
Within the statistic error, the Higgs boson mass resolution maintains at the same level if the transverse cell size is enlarged from 5~mm to 10~mm. 
When the cell size is further enlarged to 20~mm, a $\sim$5\% relative degradation is found.
Based on these observations, we recommend the ECAL cell size should be equal or smaller to 10~mm.

\begin{table}[htbp]
\centering
\caption{\label{tab:3} Resolution of reconstructed Higgs boson mass through $\nu\nu Higgs, Higgs\to gluons$ events with different cell size at CEPC\_v1 geometry.}
\smallskip
\begin{tabular}{|c|c|c|}
\hline
Silicon sensor cell size & Higgs boson mass resolution (Statistic error only) \\
\hline
5 mm  & 3.74 $\pm$ 0.02 \% \\
10 mm & 3.75 $\pm$ 0.02 \% \\
20 mm  & 3.93 $\pm$ 0.02 \% \\
\hline
\end{tabular}
\end{table}

\section{Conclusion}
\label{sec:6}
The ECAL optimization is crucial for the CEPC detector design. 
Starting from the benchmark detector geometry at CEPC Pre-CDR study (CEPC\_v1), we explore the physics performance at different ECAL geometries using Geant4 simulation and Arbor reconstruction. 
Based on the study results, we recommend a set of optimized ECAL geometry parameters. 

The local ECAL geometry is determined by the longitudinal structure and the transverse cell size. In the original design at CEPC\_v1 geometry, the ECAL has a total absorber thickness of 84 mm, divided into 30 longitudinal layers. 
Each longitudinal layer consists of one 500-micrometer silicon sensor layer, which is segmented into 5~mm readout cells (square shape). 
In this manuscript, we focus on these local ECAL geometry parameters: total absorber thickness, longitudinal layer number, silicon sensor thickness, and transverse cell size.   

The longitudinal structure is essential for the intrinsic photon energy resolution, which is characterized by single photon energy resolution and the mass resolution of Higgs boson via di-photon final states. 
Using a simplified, defect-free ECAL geometry, we find that the single photon energy resolution reaches 15.9\%/$\sqrt{E}\oplus$0.95\% (Figure~\ref{fig:8}), agrees with the CALICE test beam result~\cite{CALICE}.
A relative mass resolution of 1.64\% is achieved for the Higgs boson mass measured from di-photon final states channel, about 20\% better than the result at CEPC v$\_$1 geometry. 
Using $Higgs\to\gamma\gamma$ as the benchmark, we scan the tungsten absorber thickness. 
Our study shows that 84 mm total tungsten thickness is optimized for this benchmark measurement.

We also scan the photon energy resolution at different numbers of longitudinal layers. 
Not surprisingly, reduced number of layers leads to worse performance. 
However, thicker silicon sensor layers can be used to compensate this degradation.
Comparing the original design (30 longitudinal layers, each equipped with 500~$\mu m$ thick silicon sensor layer) and the optimized design (25 longitudinal layers and 1 mm thick silicon layer), we observe the same level performance at single photon samples.
The layer number could be further reduced to 20 layers, if thicker silicon sensor (1.5 mm) could be adopted.
This conclusion is cross-checked, and consistent with the analysis at physics benchmark of $\nu\nu Higgs, Higgs\to gluons$ events. 

The transverse cell size of ECAL dominates the shower separation performance, which is fundamental for the PFA performance. 
Using di-photon samples, we find that Arbor can efficiently separate particles as far as their impact position is distanced larger than twice of the cell size (for cell size larger than 4 mm $\times$ 4 mm and smaller than the Moliere radius). 
We study the chances of photon overlapped with other particle showers at the $Z\rightarrow\tau^+\tau^-$ samples at 91.2 GeV center of mass energy. 
Our analysis shows that the photon overlapping chance maintains at a low level for cell size smaller than 10 mm (roughly 2\%); however, increasing the cell size from 10 mm to 20 mm, the overlapping chance will also be increased by one order of magnitudes. 
Meanwhile, at $\nu\nu Higgs, Higgs\to gluons$ events sample, we observe a mass resolution of 3.74\%/3.75\%/3.93\%, corresponding to 5~mm/10~mm/20~mm ECAL transverse cell size. 
Therefore, considering the physics requirement of the precise EW measurements, we recommend the cell size to be 10 mm or smaller.

The recommended ECAL geometry parameter is therefore summarized in Table 4.
This set up fulfills the physics requirement for CEPC Higgs boson and EW measurements.
Giving similar collision environments and physics objectives, we believe this set up is also a reasonable starting point for the ECAL optimization for other electron
positron Higgs/Z factories.

\begin{table}[htbp]
\centering
\caption{\label{tab:4} Optimized ECAL local geometry parameters.}
\smallskip
\begin{tabular}{|c|c|}
\hline
Layer number  & 25 \\
\hline
Silicon wafer thickness & 1 mm \\
\hline
Tungsten plate thickness & 3.36 mm \\
\hline
Transverse cell size & 5 - 10 mm \\
\hline
\end{tabular}
\end{table}

\acknowledgments
The author thanks Gang Li and Xin Mo for the physics event generator files. 
We appreciate Vincent Boudry, Vladislav Balagura, and Henri Videau, at LLR, Ecole Polytechnique, for their insights to high granularity calorimeter and discussions.
This study is supported by National Key Program for S\&T Research and Development (Grant No.: 2016YFA0400400), National Natural Science Foundation of China (No.: 11675202 and No.: 11675196), and the CAS Hundred Talent Program (Y3515540U1)

\end{document}